\documentclass[sigconf]{acmart}

\usepackage{makecell}
\usepackage{multirow}
\usepackage{soul,color}
\soulregister\cite7
\soulregister\ref7
\soulregister\pageref7
\soulregister\figurename7
\soulregister\subsection7
\soulregister\subsubsection7
\soulregister\tablename7
\soulregister\systemname7
\soulregister\bfseries7
\soulregister\textbf7
\soulregister\oursystem7
\AtBeginDocument{%
  }

\copyrightyear{2025}
\acmYear{2025}
\setcopyright{acmlicensed}
\acmConference[UIST '25]{The 38th Annual ACM Symposium on User Interface Software and Technology}{September 28-October 1, 2025}{Busan, Republic of Korea}
\acmBooktitle{The 38th Annual ACM Symposium on User Interface Software and Technology (UIST '25), September 28-October 1, 2025, Busan, Republic of Korea}
\acmDOI{10.1145/3746059.3747626}
\acmISBN{979-8-4007-2037-6/2025/09}

\usepackage{xspace}
\newcommand{\oursystem}{NoteIt\xspace}

\definecolor{lightblue}{RGB}{66,133,244}
\usepackage{amsmath}
\usepackage{enumitem}
\usepackage{subfigure}
\newcommand{\etal}{et al.\xspace}

\newcommand{\eg}{\emph{e.g.,}\xspace}

\newcommand {\dquote}[1]{``{#1}''}

\begin{document}

\title{\oursystem: A System Converting Instructional Videos to Interactable \\ Notes Through Multimodal Video Understanding}

\author{Running Zhao}
\authornote{Both authors contributed equally to this research.}
\email{rnzhao@connect.hku.hk}
\affiliation{%
  \institution{The University of Hong Kong}
  \city{Hong Kong}
  \country{China}
}

\author{Zhihan Jiang}
\authornotemark[1]
\email{zhjiang@connect.hku.hk}
\affiliation{%
  \institution{The University of Hong Kong}
  \city{Hong Kong}
  \country{China}
}

\author{Xinchen Zhang}
\email{u3008407@connect.hku.hk}
\affiliation{%
  \institution{The University of Hong Kong}
  \city{Hong Kong}
  \country{China}
}

\author{Chirui Chang}
\email{u3010225@connect.hku.hk}
\affiliation{%
  \institution{The University of Hong Kong}
  \city{Hong Kong}
  \country{China}
}

\author{Handi Chen}
\email{hdchen@connect.hku.hk}
\affiliation{%
  \institution{The University of Hong Kong}
  \city{Hong Kong}
  \country{China}
}

\author{Weipeng Deng}
\email{wpdeng@eee.hku.hk}
\affiliation{%
  \institution{The University of Hong Kong}
  \city{Hong Kong}
  \country{China}
}

\author{Luyao Jin}
\email{lyjin@link.cuhk.edu.hk}
\affiliation{%
  \institution{The Chinese University of Hong Kong}
  \city{Hong Kong}
  \country{China}
}

\author{Xiaojuan Qi}
\email{xjqi@eee.hku.hk}
\affiliation{%
  \institution{The University of Hong Kong}
  \city{Hong Kong}
  \country{China}
}

\author{Xun Qian}
\authornote{Corresponding author.}
\email{me@xun-qian.com }
\affiliation{%
  \institution{Google}
  \city{Mountain View}
  \state{CA}
  \country{USA}
}

\author{Edith C.H. Ngai}
\email{chngai@eee.hku.hk}
\affiliation{%
  \institution{The University of Hong Kong}
  \city{Hong Kong}
  \country{China}
}
\authornotemark[2]

\renewcommand{\shortauthors}{R. Zhao, Z. Jiang, X. Zhang, C. Chang, H. Chen, W. Deng, L. Jin, X. Qi, X. Qian, C.H. Ngai}
\begin{abstract}
 
Users often take notes for instructional videos to access key knowledge later without revisiting long videos. Automated note generation tools enable users to obtain informative notes efficiently.
However, notes generated by existing research or off-the-shelf tools fail to preserve the information conveyed in the original videos comprehensively, nor can they satisfy users' expectations for diverse presentation formats and interactive features when using notes digitally.
In this work, we present \oursystem, a system, which automatically converts instructional videos to interactable notes using a novel pipeline that faithfully extracts hierarchical structure and multimodal key information from videos. 
With \oursystem's interface, users can interact with the system to further customize the content and presentation formats of the notes according to their preferences.
We conducted both a technical evaluation and a comparison user study (N=36). The solid performance in objective metrics and the positive user feedback demonstrated the effectiveness of the pipeline and the overall usability of \oursystem. 
Project website: \href{https://zhaorunning.github.io/NoteIt/}{\color{lightblue}https://zhaorunning.github.io/NoteIt/}.

\end{abstract}

\ccsdesc[500]{Human-centered computing~Interactive systems and tools}
\keywords{note generation, multimodal learning, video understanding, multimodal large language model}

\begin{teaserfigure}
  \includegraphics[width=1\textwidth]{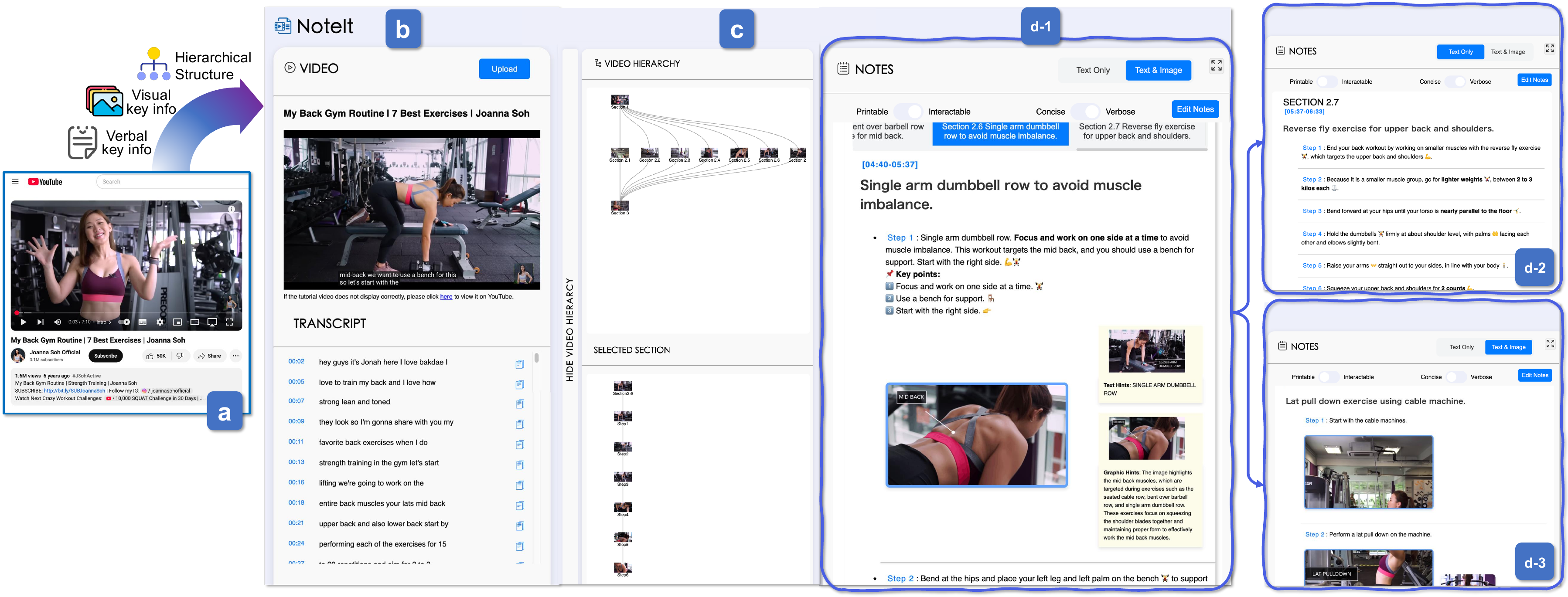}
  \caption{Overview of \oursystem. (a) A user drags an instructional video from an online resource into \oursystem's web interface. (b) \oursystem processes the video in the back end and generates interactive notes, which are displayed within the same interface. (c) The video hierarchy shows the chapter-level and step-level structures of the video, and corresponding notes are visualized as a navigable graph for quick access. (d) Users can customize the note representation according to their preferences and needs: (d-1) An interactive version includes detailed workout steps with GIFs to support direct follow-along. (d-2) A printable version enables users to bring a simplified note to the gym. (d-3) For users already familiar with the workout, a concise version highlights only the key steps emphasized by the video creator.}
  \Description{}
  \label{fig:teaser}
\end{teaserfigure}


\maketitle

\section{Introduction}
Instructional videos are visual media designed to demonstrate how to perform specific tasks step by step, covering a wide range of topics, including repair, fitness, cooking, handcrafts, first aid, and many more. In these videos, domain experts share detailed explanations and invaluable experience through various forms of representation (\eg narration, demonstrations, and visual highlights), making instructional videos a widely popular medium for learning physical tasks \cite{BeyondInstructions,MixT}.
Taking notes facilitates knowledge construction from instructional videos by retaining essential information \cite{VideoMix,VideoSticker,hefter2024note}, enabling learners to quickly review knowledge without the need to rewatch the entire video. 
However, manual note-taking can be time-consuming and cognitively demanding, prompting both commercial products \cite{notebooklm,notegpt} and Human-Computer Interaction (HCI) researchers \cite{educationnote1,educationnote2,visualtranscripts} to explore automatic note generation methods. Despite these efforts, generating high-quality notes automatically from video content remains a significant challenge.

Instructional videos often exhibit a flexible hierarchical structure as the task taught in the video may be organized either sequentially or in parallel \cite{RecipeScape,Subgoal,flexstructure,flexstructure2}. For example, in a first aid task, steps such as opening the airway, checking breathing, and chest compressions must follow a fixed sequence, whereas chest-strengthening exercises, such as push-ups, bench presses, and dumbbell flyes, can be undertaken in any order.
This structure becomes increasingly complex when a task involves both sequential and parallel components, further complicating the interpretation of the overall task flow. For example, when assembling a cabinet, components such as drawers, doors, and frames can be assembled in parallel, as there is no strict requirement for their order; while, the assembly of each individual component must follow a specific sequence, and there is also a defined order for integrating these components into the final structure.
Due to the inherently linear nature of video presentation, it is challenging to accurately model or summarize the complex underlying hierarchical structure of such physical tasks by simply analyzing the video timeline.

Furthermore, as physical tasks often involve critical details, content creators would employ various verbal and visual cues to better convey key information \cite{chi2013democut}, such as verbal emphasis, text overlays, graphic and diagram annotations, and close-up shots.
While this multimodal presentation aids human comprehension, it poses challenges for automated systems attempting to accurately capture all key information intended by the content creator. For instance, in first-aid videos, a critical instruction like \dquote{only do this if the person is unconscious} may appear briefly as an on-screen text overlay, requiring both contextual and visual understanding. Similarly, a repair video might visually highlight a small component with a red circle while simultaneously providing a verbal warning about its fragility. The multimodal nature of instructional videos highlights the need for note-generation systems that can jointly reason across visual, verbal, and textual content of videos.

Multimodal large language models (MLLMs) have demonstrated impressive capabilities in video understanding and reasoning tasks \cite{zhang2023video, zhu2023minigpt, chen2023video}, and their zero-shot generalization enables them to handle a wide range of video inputs and objectives without task-specific supervision \cite{SocraticModels}. 
However, directly prompting MLLMs to summarize instructional videos falls short of capturing the nuanced structure and critical content inherent in these videos. MLLMs typically lack structure awareness, generating flat summaries that overlook step-by-step logic and the presence of parallel and optional steps. Moreover, MLLMs may struggle to align visual and verbal modalities, failing to detect key visual-only cues like silent brush strokes in a makeup tutorial or tool orientation in a first-aid demonstration. These limitations highlight the need for a dedicated pipeline designed to handle the structural, temporal, and multimodal complexities of instructional video content.

Another important challenge lies in the presentation of the generated notes. In practice, users vary widely in their preferences for consuming instructional content. 
For example, novice users learning cooking or first aid may prefer detailed, step-by-step instructions, while more experienced users following a fitness routine may favor concise summaries. 
Moreover, some users prefer text-only notes for quick reading, whereas others find text-image formats more immersive and effective, especially in visually oriented domains like makeup tutorials. Preferences regarding engagement mode also differ among users; some favor static printable notes for immediate consultation, whereas others prefer interactive notes that reveal content via collapsible or hover-to-reveal elements to ease cognitive load, as in device repair guides that conceal detailed steps until expanded.
However, current MLLMs typically generate unstructured, free-form text, lacking the flexibility to support multiple presentation styles, levels of detail, or engagement modes (e.g., printable checklists, concise text summaries, or interactive, expandable step-by-step, text-image guides). This limitation hinders adaptation to diverse user needs and usage scenarios.
To address this gap, we endeavor to develop an interactive system that transforms instructional videos into notes in a user-preferred format, enabling tailored and faithful content delivery aligned with individual preferences and task-specific needs.

In light of these challenges and opportunities, we present \oursystem (Figure \ref{fig:teaser}), a novel system that faithfully converts instructional videos into interactive notes. \oursystem is designed to handle videos with flexible hierarchical content structures and diverse presentation modalities, generating notes that maintain structural consistency and comprehensively capture key information.
Users can select the output format based on their preferences and usage scenarios.
Powered by MLLMs, \oursystem introduces a modular pipeline that first parses the input video to extract its hierarchical structure and key content presented in visual and verbal formats. This parsed information is then transformed into a well-defined note scheme and rendered through a user-friendly, interactive interface.
In summary, our work makes the following contributions:

\begin{itemize}[leftmargin=*]
    \item A design space for generating interactive notes that maintain structural consistency with the instructional videos and incorporate key verbal and visual information across diverse instructional videos.
    \item An end-to-end pipeline that processes instructional videos to extract hierarchical structures and multimodal key content, and systematically maps them into structured notes.
    \item An interactive user interface that enables users to upload instructional videos and explore the generated notes with diverse presentation modalities, verbosity levels, and engagement modes.
\end{itemize}

\section{Related Work}

\subsection{Mixed-media Tutorials for Instructional Videos}
Mixed-media tutorials are essential tools to facilitate users in learning skills from instructional videos by engaging multiple auxiliary components (including text descriptions, thumbnails, and timestamps). The HCI community has devoted considerable effort to advancing these tools for enhancing instructional video experiences \cite{OnPause,nawhal2019videowhiz,MixT,pavel2014video,Crowdsourcinghowto}. Their focus has gradually shifted from well-structured videos, such as software or smartphone usage \cite{MixT,EverTutor}, and educational videos \cite{pavel2014video}, to instructional videos on physical tasks, such as cooking, handcraft, makeup, and so on. 
To enhance the learning of physical tasks, various mixed-media tools are designed by integrating step-related information (step description and dependencies) \cite{nawhal2019videowhiz,RecipeScape,mixmedia_makeup,Soloist} and visual effects \cite{chi2013democut,isualizinginstructions,VideoMix}.
For example, VideoWhiz \cite{nawhal2019videowhiz} presented a non-linear browsing strategy to benefit cooking video overview. Truong et al. \cite{mixmedia_makeup} proposed a multi-modal approach to extract fine- and coarse-level steps to ease make-up video navigation. Researchers in \cite{isualizinginstructions} explored the potential of visual cues to help learning from physical training instructional videos. Based on them, an AI-assisted framework, TutoAI \cite{tutoai}, is proposed to create mixed-media tutorials from instructional videos.

Beyond mixed-media tutorials, note-taking remains a traditional yet powerful medium, facilitating skill acquisition from instructional videos \cite{VideoMix,VideoSticker,hefter2024note}. Notes can further augment mixed-media tutorials by providing flexible formats without relying on video itself and incorporating comprehensive information. This paper follows the mixed-media tutorial and instructional video paradigm as discussed above, and further proposes an advanced note-generation system.
NoteIt not only enables auxiliary components retrieval (see above) and video navigation within mixed-media tutorials, but also introduces novel capabilities for extracting and visualizing hierarchical structures and multimodal key information. Existing mixed-media tutorials only extract the coarse-grained steps, while NoteIt extracts and represents the hierarchical structures at both chapter and step levels, providing users with a clearer and more effective learning pathway. Moreover, NoteIt comprehensively captures verbal and visual key information emphasized by creators, rather than merely gathering thumbnails and step descriptions as tutorials do. In contrast to crowd-powered tutorials \cite{RecipeScape,Crowdsourcinghowto,nawhal2019videowhiz,pavel2014video,Subgoal}, NoteIt automatically generates notes of consistent quality; although crowd-sourcing offers lightweight solutions with community support, the computational cost of deploying the more powerful NoteIt is worthwhile.

\begin{figure*}[h]
    \centering
    \includegraphics[width=1\textwidth]{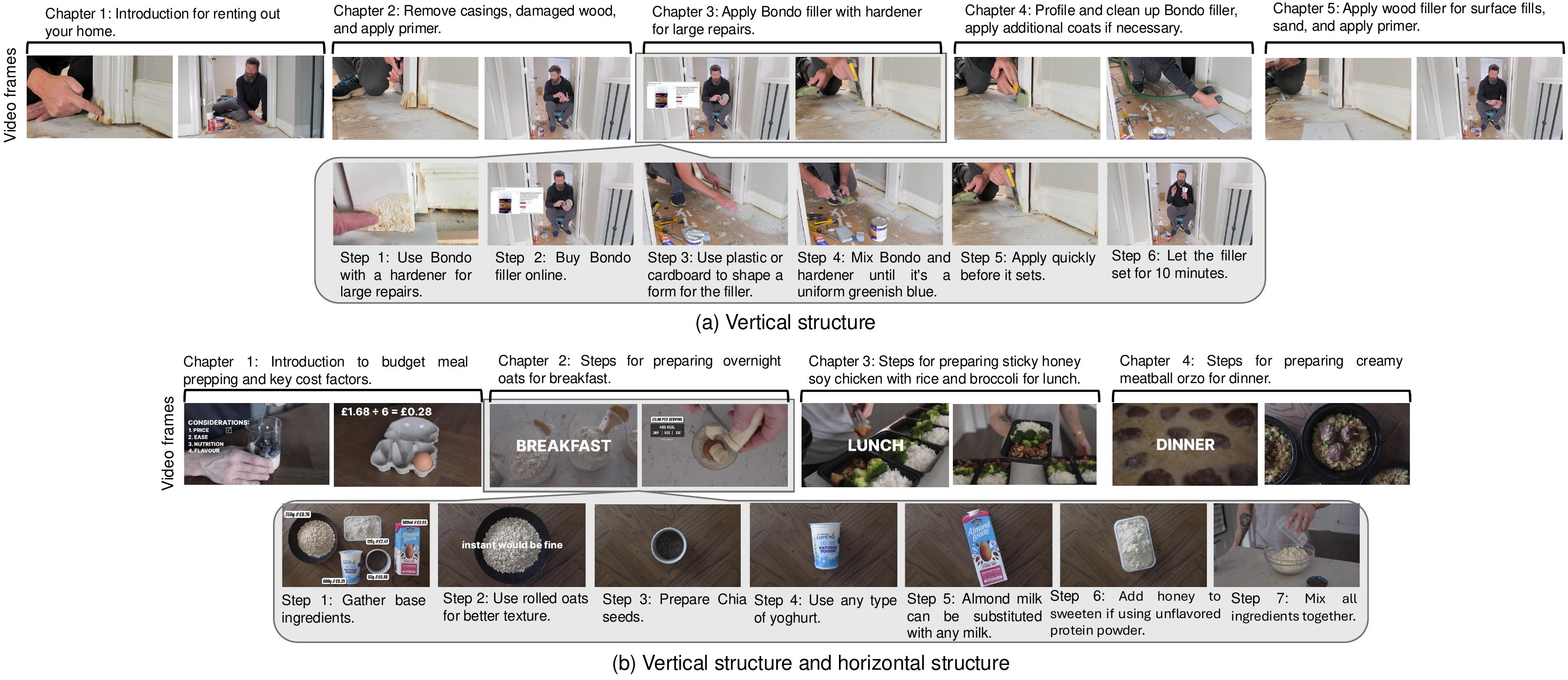}
    \caption{Example of hierarchical structures in the video repairing a damaged door and the video cooking meal on a budget. (a) The video follows a vertical structure across all chapters as they are presented in a sequential manner. The steps within Chapter 2 are carried sequentially, and thus, they are also a vertical structure. (b) In this video, Chapter 2 through Chapter 4 are parallel, as they can be performed in any order. Similarly, within Chapter 2, Step 2 through Step 6 are also parallel steps.}
    \label{fig:structure}
\end{figure*}

\subsection{Note-taking and Note Generation}
Note-taking serves as a means of externally storing knowledge for future reference \cite{Notetakingfunctions}. 
Researchers in HCI community have explored various note-taking tools to enhance efficiency and experience \cite{Scraps,texSketch,HyNote,livenotes,vrnotes}; likewise, note-taking has been proven to be beneficial in constructing knowledge from instructional videos \cite{VideoMix,VideoSticker,hefter2024note}.
Correspondingly, some systems and tools have been investigated to assist with note-taking for instructional videos \cite{Gazenotes,mrnotes,VideoSticker}.

Automatic note generation is an efficient way to convert key knowledge into informative notes, alleviating the drawbacks of manual note-taking, such as frequent playback, cognitive distraction, and the potential omission of critical information \cite{Gazenotes,educationnote1}. A series of works focuses on lectures or educational videos.
For example, Xu \etal \cite{educationnote1} and Xu \etal \cite{educationnote2} extracted visual entities in the video slides and speech transcripts to generate lecture notes, and Shin \etal \cite{visualtranscripts} proposed Visual Transcripts that generate lecture notes from both the visual and audio content of blackboard-style lecture videos. 
However, unlike lecture videos that follow a fixed structure with a predefined outline or progression and convey content through structural slides, instructional videos on physical tasks have flexible structures and complex presentation formats, making them more challenging to process.
Powered by LLM, existing commercial off-the-shelf tools, such as NotebookLM \cite{notebooklm} and NoteGPT \cite{notegpt}, are capable of tackling instructional videos on physical tasks to generate notes. However, these tools generate only plain-text summarization. Specifically, they mechanically list the step summaries without explicating step hierarchical organizations and inter-step relations, hindering users from systematically grasping the workflow of the physical task. Furthermore, their text summaries inadequately convey the visual information highlighted by creators, leading to the omission of key information. In addition, their fixed text presentation format limits the users' selections, failing to accommodate diverse preferences.

NoteIt faithfully captures chapter- and step-level hierarchies and displays their relationships, while comprehensively conveying key information in visual and verbal modalities. NoteIt also enables the users to select the presentation modality, content verbosity, and engagement mode according to their preferences.

\subsection{Instructional Video Understanding}

Recent advancements in Vision-Language Models (VLMs) have significantly enhanced video understanding by integrating semantic interpretation with temporal reasoning. While models like BLIP-2 \cite{li2023blip}, Video-LLaMA \cite{zhang2023video}, and MiniGPT-4 \cite{zhu2023minigpt} have shown promise in video captioning and summarization, they remain limited in capturing the long-range structure and fine-grained illustrations needed for instructional tasks.
Models such as Vid2Seq \cite{yang2023vid2seq} and Video ChatCaptioner \cite{chen2023video} extend dense captioning to detailed multi-sentence summarizations for long, complex videos, while VidIL \cite{wang2022language} and FAVD \cite{shen2023fine} enable few-shot and frame-level captioning to capture fine-grained actions. Moreover, researchers also explored the generating and captioning capability of GPT-4 \cite{chen2024sharegpt4video} and GPT-4V \cite{han2023shot2story20k, yang2024vript}, respectively, indicating significant improvement. More recently, GPT-4o demonstrates strong multimodal reasoning abilities, making it well-suited for instructional video understanding. Leveraging its powerful capabilities in multimodal understanding, NoteIt is built upon GPT-4o (GPT-4o Vision).

While these advanced MLLMs demonstrate the potential in video understanding, directly prompting them to summarize video content with step descriptions remains insufficient for representing hierarchical structure relationships and conveying comprehensive key information that creators highlight. \oursystem integrates MLLMs into a novel pipeline for instructional video understanding, going beyond simple video summarization to capture and represent hierarchical structure and multimodal key information that supports more effective comprehension for instructional videos.

\section{Design Space}
\label{designspace}

To design an automated video note-taking system, we analyzed the selected representative instructional videos and corresponding learner notes, focusing on instructional videos on physical tasks. Based on the analysis, we derive a set of design goals for generating interactive notes from instructional videos. As it is impractical to cover all possible videos and notes for analysis, deriving common characteristics from representative videos and notes provides a reasonable basis for generalizable design goals. We employed a purposive sampling to select representative instructional videos and notes, like existing HCI works \cite{tutoai,chi2013democut,HowToCut}. Thus, the automatic note generation system grounded in these design goals can generalize across diverse instructional videos and tasks.

\begin{figure*}[h]
    \centering
    \includegraphics[width=1\textwidth]{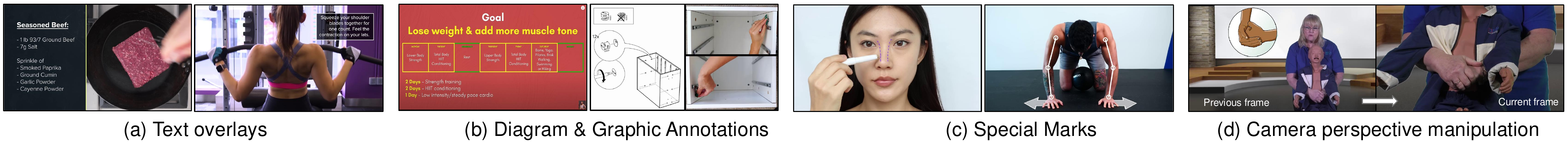}
    \caption{Example of visual key information: (a) text overlays of cooking beef and lat pull-down exercise, (b) diagram \& graphic annotations of fitness plan and drawer assembly, (c) special marks of makeup and fitness routine, (d) camera perspective manipulation from a wide view presenting arm and body position to a close-up capturing detailed hand posture for a quick upward thrust.}
    \label{fig:keyinfo}
\end{figure*}

\subsection{Characteristics of Instructional Videos}
While note-taking can be applied to any type of video, it is not always necessary or meaningful across all contexts. Instructional videos, particularly those that teach physical tasks, present structural and goal-oriented knowledge, and note-taking in this context serves a useful function. 
To gain generalizable insight, we conduct a purposive sampling to analyze 80 public YouTube videos across 8 common categories: cooking, repair, assembly, handcraft, first aid, makeup, fitness, and device instruction. These categories cover a broad range of everyday skill-learning tasks. The selection criteria are as follows: (1) each video is from a different creator; (2) each focuses on a distinct task within its category (e.g, repairing a door jamb vs. a fridge); and (3) videos vary in duration. 
These selection criteria allow us to capture a wide variety of videos, ensuring representativeness.
We asked 6 raters to annotate the video structure and key content in visual and verbal formats, and each video was annotated by 2 raters and checked by 2 raters to reduce bias.

\subsubsection{Vertical and horizontal structures} 
\label{verhorstru}
Among the 80 analyzed videos, 63 videos exhibit a combination of sequential and parallel structures, 16 videos feature only sequential structures, and 1 video contains exclusively a parallel structure (see details in Appendix). Given the results, we found that instructional videos typically have a vertical structure for sequential content presentation or a horizontal structure for parallel or alternative content presentation. The identified generalizable patterns in terms of hierarchical structure are elaborated as follows.

\textbf{Vertical structure for sequential content presentation.}
Instructional videos typically have sequential content to present
chapter-by-chapter or step-by-step instructions. Specifically, chapters typically follow a logical progression from overall goals to implementation and, finally, to summary, while steps within each chapter elaborate detailed actions in order. We define this sequential content presentation as a \textbf{vertical structure}.
This structure aligns with natural viewing and cognitive processing flows, guiding viewers to follow the task from coarse to fine granularity. For example, as shown in Figure \ref{fig:structure}a, a video on repairing a damaged door follows a strictly vertical structure across all chapters, where the introduction (chapter 1) and every implementation step (Chapter 2 to 5) are presented sequentially. Moreover, within Chapter 2, Steps 1 through 6 are carried out sequentially to achieve the action of \emph{Apply Bondo filler with hardener for large repairs}. Prior research has also recognized the prevalence of vertical structures in instructional videos \cite{flexstructure,flexstructure2} and the importance of capturing such vertical structuring in facilitating comprehension and learning in instructional contexts \cite{Subgoal,RecipeScape,Informationalunits}.

\textbf{Horizontal structure for parallel or alternative content presentation.} 
We observed that not all parts of the task follow a strict sequential order: some chapters or steps can be performed independently, without dependency on others. Although videos must be presented linearly due to their temporal nature, these parallel or alternative chapters or steps are arranged with annotations (e.g., timestamps, textual or verbal descriptions) to indicate their alternative relationships and facilitate navigation. 
We define these chapters or steps that are parallel or alternative to each other as \textbf{horizontal structure}. 
This structure allows viewers to understand the flexibility of task execution and improves comprehension of parallel actions.
For example, as illustrated in the video on simple high-protein meal preparation on a budget (Figure \ref{fig:structure}b), Chapters 2 through 4 represent parallel processes: preparing breakfast, lunch, and dinner, which can be completed in any order. This means these three chapters adhere to a horizontal structure. Within Chapter 2, Steps 2 through 6 involve preparing different ingredients (e.g., oats, chia seeds, yogurt, milk, protein powder) and can be carried out independently. A similar horizontal structure appears in makeup videos, where different facial areas can be addressed in parallel.
This pattern of parallel or non-sequential structuring has also been identified in prior work \cite{flexstructure,flexstructure2}. Moreover, the importance of recognizing such parallel dependencies for instructional videos is highlighted by \cite{RecipeScape,Informationalunits,tutoai}.

\subsubsection{Key information presentation}
\label{keyinfo}
Out of the 80 videos analyzed, 65 videos present key visual information, and 78 videos include key verbal information (see Appendix). We can derive that instructional videos consistently highlight key information through both verbal and visual means.
Verbally, creators emphasize critical steps using voiceover narration, such as providing tips and warnings (e.g., precise temporal or quantitative details).
For example, in a cooking video, the narrator may say, \emph{``Let it cook for \underline{ten minutes}''}. 
Visually, key information is reinforced through text overlays, visual cues (e.g., graphic \& diagram annotations and special marks including circles, arrows, and lines) or camera perspective manipulation (e.g., shot transition for zoom-in or close-up), detailed as follows:
\begin{itemize}[leftmargin=*]
    \item \textbf{Text overlay} is widely used to enhance comprehension and guide viewer attention. Creators add plain text to video frames to emphasize key terms or instructions as visual anchors \cite{chi2013democut,HowToCut}. For example, in Figure \ref{fig:keyinfo}a, text overlays are used to label cooking ingredients for beef and highlight essential actions in a lat pull-down exercise. 
    \item \textbf{Graphic and diagram annotations} help clarify complex relationships and make abstract concepts more accessible \cite{VideoSticker}. These structured visual representations are either overlaid on empty areas of video frames or occupy entire frames. As shown in Figure \ref{fig:keyinfo}b, a table outlines the fitness plan for the training goals, and an instructional diagram illustrates how to lock the plastic cam into position.
    \item \textbf{Special marks} such as circles, arrows, and lines are frequently used to highlight specific elements, ensuring viewers focus on the details \cite{isualizinginstructions,chi2013democut}. Specifically, circles enclose specific parts, arrows indicate direction or causality, and lines connect related elements or delineate sections (Figure \ref{fig:keyinfo}c).
    \item \textbf{Camera perspective manipulation} is strategically used to guide viewers' attention toward key information \cite{chi2013democut,HowToCut}. Creators alternate between wide shots for context and close-ups for detail, guiding viewer focus through visual transitions. In Figure~\ref{fig:keyinfo}d, a wide view presents the arm and body position, followed by a close-up capturing the detailed hand posture for a quick upward thrust.
\end{itemize}
Recent works \cite{VideoSticker,HowToCut,chi2013democut} suggest instructional videos commonly employ similar video editing techniques to highlight key information.

\subsection{Characteristics of Notes}
\label{charnotes}
We also employed a purposive sampling of notes to select representative notes, deriving generalizable patterns. We identify the characteristics of notes from two levels. First, to understand effective and commonly accepted note-taking practices, we collected 10 highly rated note templates from the commercial note-taking software Notion, where each template was endorsed by hundreds of users through high scores and positive comments. Additionally, to incorporate instruction domain-specific patterns, we analyzed tutorial documents that accompany instructional videos, focusing on how users document instructional videos or tasks. We collected 30 tutorials from platforms including iFixit, wikiHow, Google Help, Instructables, and Allrecipes.com. This dual-layered approach enables us to derive a design space that is both representative of real-world practices and directly aligned with our system's goals. Our findings are summarized as follows. 

\begin{figure}[t]
    \centering
    \includegraphics[width=0.48\textwidth]{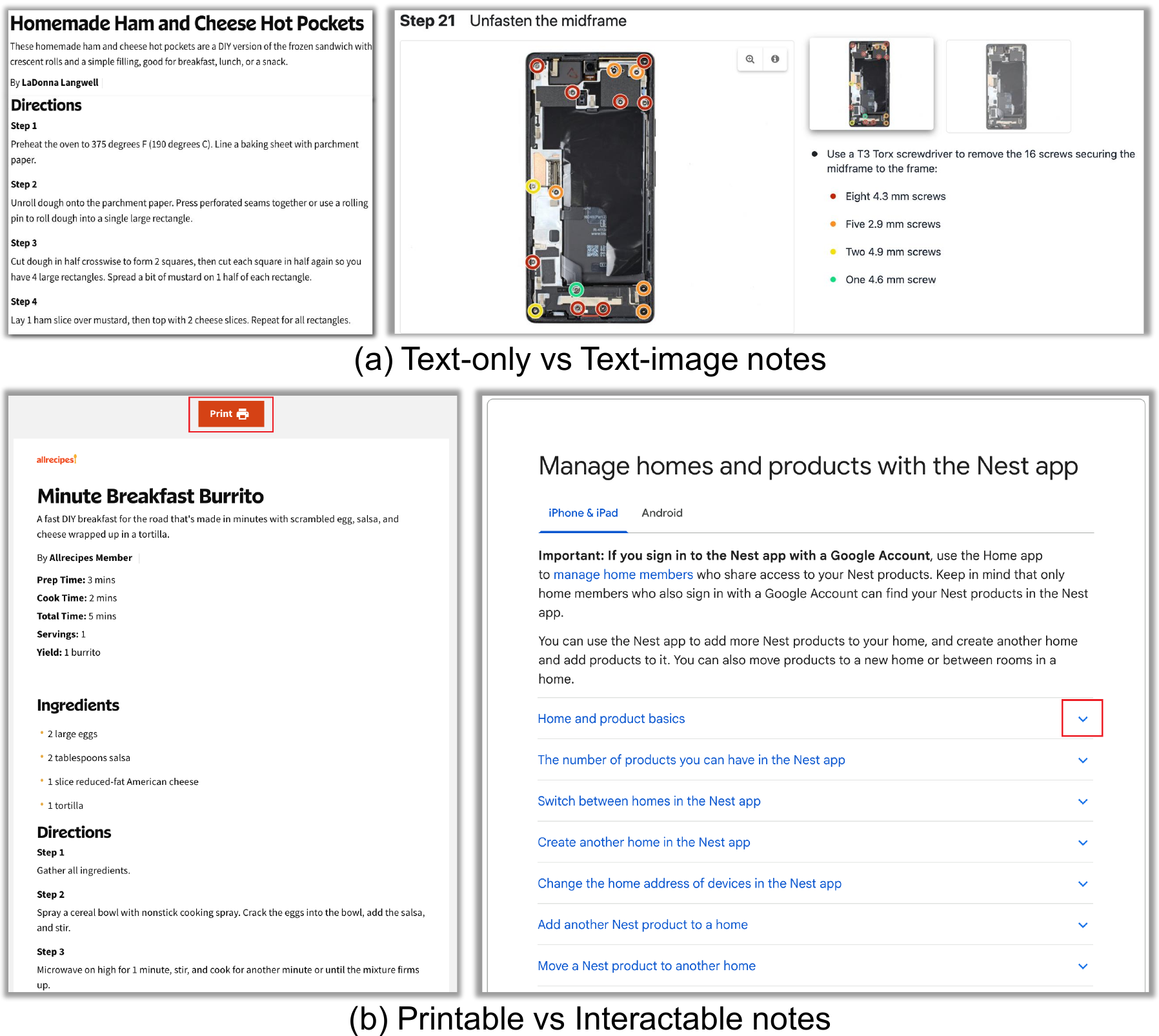}
    \caption{Examples of notes presentation modality and engagement mode: (a) text-only and text-image modalities. (b) printable and interactable modes.}
    \label{fig:noteformats}
\end{figure}

\textbf{Presentation Modality.} 
Notes vary in presentation modality, with some using text-only formats and others incorporating multimodal elements such as images or video clips. 
Text-only notes rely solely on written instructions, making them easy to scan and follow. For example, the note for making \emph{ham and cheese hot pockets} presents clear procedural steps, such as \emph{``Unroll dough onto the parchment paper''} and \emph{``Press perforated seams together or use a rolling pin to roll dough into a single large rectangle''} (Figure \ref{fig:noteformats}a). 
In contrast, multimodal notes combine text with visuals to provide additional context and support comprehension. For instance, a battery replacement guide includes an image showing the position of 16 screws alongside the instruction: \emph{``Use a T3 Torx screwdriver to remove the 16 screws securing the midframe to the frame"} (Figure \ref{fig:noteformats}a).
Beyond content complexity, users’ preferences for processing information also influence modality: text-only notes benefit those who favor textual analysis, while text-image formats support visual learners \cite{wang2024online}. This distinction aligns with user-centered design principles that emphasize adaptability to different cognitive styles.

\textbf{Content verbosity.}
Notes vary in their level of content verbosity. Some provide detailed step descriptions, including execution parameters, contextual cues, and possible variations. For example, a makeup tutorial advises: \emph{``Apply concealer with a patting motion using your ring finger, focusing on the under-eye area while avoiding pulling your delicate skin."}
In contrast, we also found that some of the notes offer concise descriptions, reducing steps to their essential components such as a device instruction notes \emph{``Adjust brightness: Settings → Display → Brightness slider."} 
In contrast, other notes adopt a concise style, distilling steps to essential actions, such as: \emph{``Adjust brightness: Settings → Display → Brightness slider."}
These differences reflect user interaction patterns, where novice users tend to prefer detailed guidance, and expert users favor concise notes that act as memory cues. Prior design work \cite{designmagic} supports this dual approach, showing that offering both comprehensive and condensed content enhances cognitive flexibility.

\textbf{Engagement mode.} 
Notes commonly follow one of two engagement modes: printable or interactable. Printable notes present all content at once in a static, linear format, making them readily accessible for direct use without user interaction.
For example, recipe notes can be printed and referenced immediately (Figure~\ref{fig:noteformats}b). 
In contrast, interactable notes adopt a dynamic format, revealing information progressively through features such as collapsible sections, hover-to-reveal terms, or clickable elements.
For example, in a device instruction notes, detailed steps remain hidden until expanded by the user (Figure \ref{fig:noteformats}b). 
These two modes align with the principle of progressive disclosure, which aims to reduce cognitive load by presenting only relevant information when needed \cite{ribeiro2020exploring,ding2020progressive}. Thus, printable and interactable notes serve distinct cognitive functions—one prioritizing accessibility, the other supporting focused engagement.

\begin{figure*}[h]
    \centering
    \includegraphics[width=1\textwidth]{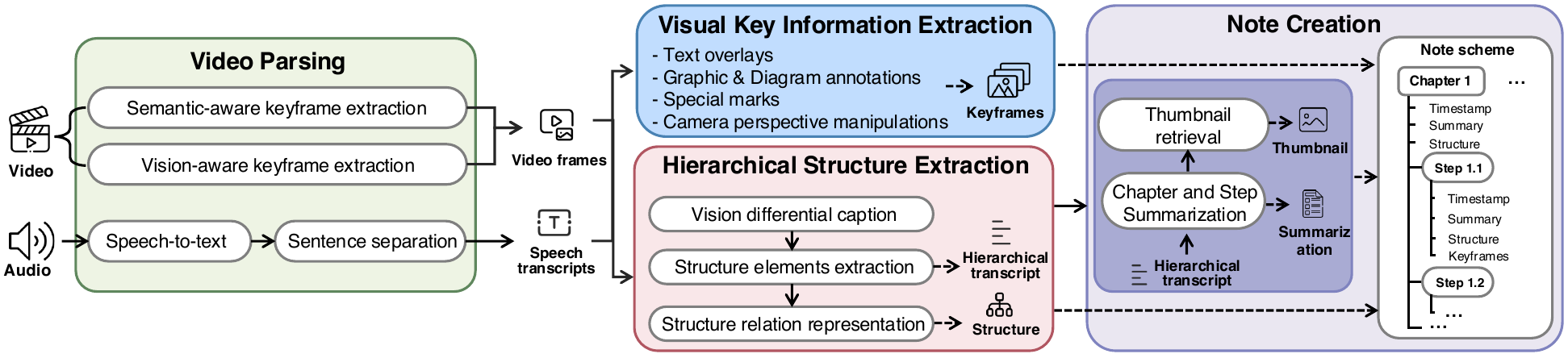}
    \caption{\oursystem's pipeline consists of four modules. \emph{Video parsing} module preprocesses video to extract video frames without redundancy and audio to extract speech transcripts. \emph{Hierarchical structure extraction} module extracts and represents structures at chapter and step levels. \emph{Visual Key information extraction} module selects key frames containing various presentations of visual information. \emph{Note creation} module completes the notes content and represents it as a designed structural scheme.}
    \label{fig:framework}
\end{figure*}

\subsection{Design Goals}

Given the representativeness of the selected instructional videos and notes, they provide empirical grounding for our design goals analysis. Based on our analysis of instructional videos and notes, we identify four key design goals to guide the development of a system that converts instructional videos into notes.

\textbf{D1. Maintain note structure consistent with the original video.} The note generated should maintain the vertical and horizontal structures of the instructional video, accurately reflecting both sequential and parallel flows (Section \ref{verhorstru}). 
For vertical structure, notes should present chapters and steps in a top-down, step-by-step format that mirrors the video's order.  
For horizontal structure, parallel or alternative chapters and steps should be shown side-by-side without crossing them with vertical structures. This structural alignment supports clarity and faithful representation of the instructional content \cite{Subgoal,RecipeScape,tutoai}.

\textbf{D2. Include both visual and verbal key information.} 
As analyzed in Section \ref{keyinfo}, since instructional videos emphasize key information both verbally and visually, the notes should capture verbal key information (e.g., tips, warnings, and precise instructions from narration) and visual key information (e.g., text overlays, graphic and diagram annotations, special marks, and camera perspective manipulations). Incorporating both modalities ensures that critical information is preserved, enabling users to follow the instructions accurately and effectively \cite{VideoSticker,chi2013democut}.

\textbf{D3. Support interaction based on user preferences.} 
As analyzed in Section \ref{charnotes}, to accommodate diverse user requirements, task familiarity, and usage scenarios, the note-generation system should offer multiple options, including modality (text-only or text-image/GIF pairs), content verbosity (concise or detailed), and engagement mode (printable or interactive). This flexibility ensures that notes align with individual cognitive styles and practical needs.

\textbf{D4. Scale across diverse instructional videos.}
Given the varying characteristics of instructional videos analyzed in Section \ref{designspace} across categories, structures, and content types, the system should be scalable and adaptable without relying on handcrafted rules. It should automatically generate notes tailored to each video, enabling broad applicability and meeting the needs of diverse users.

As these distilled design goals capture the common patterns, the resultant note generation system is established on principles that are agnostic to video content and domain, generalizing across diverse instructional videos beyond the selected videos.

\section{\oursystem}

Following the set of design goals, we developed \oursystem, a system that automatically converts instructional videos to interactable notes. The input is instructional videos with flexible hierarchical structure and diverse presentation of key information. With \oursystem's UI, users can interact with the notes to further customize the content (levels of content verbosity) and presentation format (presentation modality and engagement mode) of notes according to their preferences. As shown in Figure \ref{fig:framework}, our system contains five modules: (1) \emph{video parsing} that preprocesses video to extract video frames without redundancy and audio to extract speech transcripts; (2) \emph{hierarchical structure extraction} to extract and represent structures in chapter and step levels; (3) \emph{key information extraction} to select key frames with various presentations of visual information; (4) \emph{note creation} to complete the notes content and represent it as a designed structural scheme; (5) \emph{interactable UI} for user customization.

\begin{figure*}[t]
    \centering
    \includegraphics[width=0.87\textwidth]{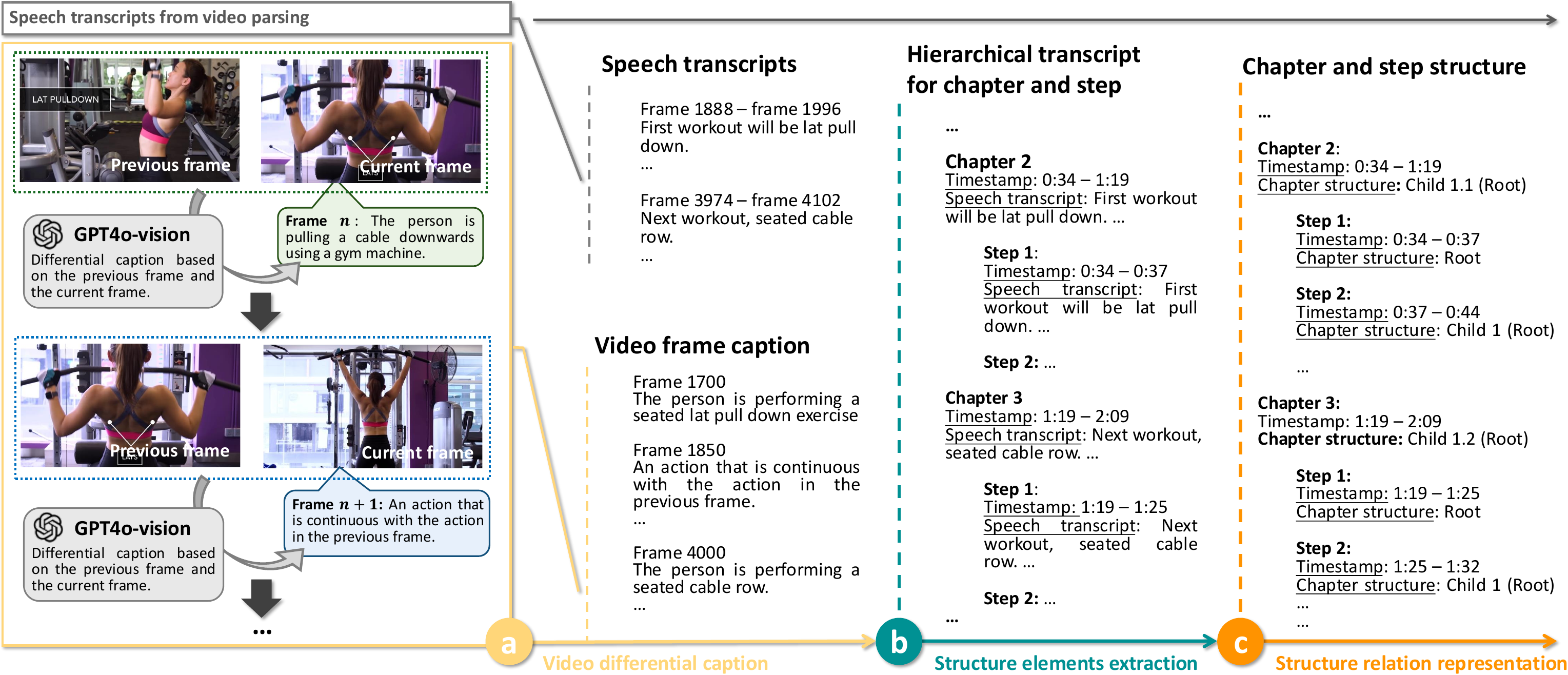}
    \caption{\oursystem hierarchical structure extraction pipeline. (a) Vision differential caption method generates video frame caption from the filtered video frames. (b) The video frame caption, along with speech transcripts extracted from video parsing, are transformed into the hierarchical transcript. (c) The hierarchical transcript is transformed into chapter and step structure.}
    \label{fig:hierarchy}
\end{figure*}

\subsection{Video Parsing}
We first parse the video to obtain the basic elements (images and text) from the raw instructional video. The output of filtered video frames and speech transcripts are passed to the following visual key information and hierarchical structure extraction modules.

To reduce the redundancy of the original video with a high frame rate, we employ a dual approach leveraging CLIP for semantic-aware keyframe extraction and DINO for vision-aware keyframe extraction. 
We leverage CLIP \cite{clip}, a powerful model to understand visual content in a semantically meaningful way, to extract frames representing the core content of the video. Specifically, we compute the visual CLIP embeddings for each frame and calculate the cosine similarity between two consecutive embeddings \cite{jiang2025dietglance}. We then compare the similarity score with the threshold to filter out the semantic-aware keyframes. But a purely semantic-aware method may miss some vision critical frames. To cope with that, we incorporate the vision-aware method to extract the frames with visually distinct moments, even if their semantic relevance is lower. DINO \cite{DINOv2} is employed due to its proficiency in capturing visual distinctiveness and structural variations in images. Similarly, we process each frame through DINO to obtain embeddings and filter the redundancy based on the cosine similarity. Finally, we identify the intersection of these two filtered sets as keyframes. 
For the speech signal accompanying the original video, we use the speech-to-text model, Whisper \cite{whisper}, to transcribe the speech into text, and Whisper provides sentence segmentation, which we leverage to divide the text into coherent sentence-level units for further analysis.

CLIP and DINO are pre-trained models that excel in semantic and visual generalization, while Whisper also demonstrates its robustness in varied conditions, making the parsing method scalable to video with different types and content (\textbf{D4}).

\subsection{Hierarchical Structure Extraction}
This module aims to extract the hierarchical structure in a chapter level and step level, and represent them as a directed acyclic graph (DAG), respectively, to clearly show their vertical and horizontal structure (\textbf{D1}). The extraction of hierarchical structures facilitates effective learning from physical tasks \cite{Subgoal,NiCEBook}. To achieve this goal, our module sequentially performs vision differential caption, structure elements extraction, and structure relation representation. This module is built on the GPT-4o with robust scalability (\textbf{D4}).

First, we leverage a vision differential caption method based on GPT4o-vision (Figure \ref{fig:hierarchy}a) to generate high-quality captions containing temporal relations. The intuitive way is to prompt GPT4o-vision to caption each frame or concatenate all the frames into a large image for caption, but the generated caption struggles with the correct temporal relation or the details. In contrast, we feed the previous key frame and current key frame into GPT4o-vision. Then, we use a prompt to guide GPT4o-vision to \emph{compare the current frame with the previous one and judge whether a change between them} and generate the caption according to the observation: \emph{if there present changes between two input frames, describe the changes; otherwise, claim the current frame is continuous with the previous one.} The differential caption process continues until all frames have been captioned. 

Second, based on the generated caption of key frame and the sentence level transcript with time stamps, we use GPT-4o \cite{GPT4o} to cluster the content into chapters and extract steps from each chapter to construct the basic structure elements of instructional videos (Figure \ref{fig:hierarchy}b). To form the chapter-level structure, we prompt GPT-4o to \emph{cluster all frame indices into multiple frame index sets (chapters) with time stamp based on the input captions and speech transcript}, denoted as $C=\{c_i (t_s, t_e)\}$, where $t_s, t_e$ represent start and end time. The corresponding frame caption and speech transcript within the chapter start and end time are extracted as the chapter content. For each chapter, we prompt GPT-4o to \emph{summarize the content into key steps with time stamps based on the chapter content}, represented as $S=\{s_{ij}(t_s, t_e)\}$, where $t_s, t_e$ represent start and end time. Correspondingly, the frame caption and the speech transcript within the step start and end time are also extracted as the step content. The chapter content and step content comprise the hierarchical transcript, the hierarchical structure with corresponding transcripts.

Finally, we represent the vertical and horizontal structure of chapters and steps (Figure \ref{fig:hierarchy}c). To achieve this, we design the Chain-of-Thought reasoning steps that break the hierarchical structure extraction tasks (chapter and step levels) into two sub-tasks, including relation extraction and representation based on DAG, and guide GTP-4o to execute them step by step. We prompt GPT-4o to \emph{identify the logical relation between chapters (or steps) based on the corresponding content and categorize them as sequential or parallel and alternative relation}. Then, it iteratively constructs a DAG to represent the identified relation between chapters (or steps). Formally, in the prompt template, we define the DAG as $G=(V,E)$, where nodes $V=\{v_1,...,v_n\}$ represent the structure elements (chapters or steps) and directed edges $(v_m, v_n)\in E$, where $E \subseteq V\times V$ (for every sequence of edge $(v_1, v_2),...,(v_{k-1}, v_k)$, it must hold that $v_1 \neq v_k$), encode the sequential relation as the predecessor to successor and the parallel or alternative relation as multiple successors from the same predecessors. With the DAG-based representation, the vertical and horizontal structures of chapters and steps are extracted to faithfully reflect the hierarchical structure in the original video.

\subsection{Visual Key Information Extraction}
This module aims to extract frames representing visual key information from filtered images (\textbf{D2}). To achieve this, the module presents an agentic workflow to extract the static key frames (text overlays, graphic and diagram annotations, and special marks) and leverage a suit of vision and language models to extract the dynamic key frames (camera perspective manipulation), respectively.

\subsubsection{Static key frames}
As analyzed for visual key information, text overlays, graphic and diagram annotations, and special marks are static information existing in the corresponding key frames. Therefore, they can be detected by directly inspecting whether video frames contain such elements. Since these visual key information have distinct characteristics and also vary across different videos, it is challenging to establish criteria for extracting such visual key information from each video. Therefore, we propose an agentic workflow to adaptively derive key frames containing static information from various videos (\textbf{D4}), where GPT-4o acts as a planner to formulate high-level sub-tasks given the query task and GPT4o-vision acts as an executor to perform the sub-tasks that humans further refine. The planner operates as a reasoning module that interprets user-specified tasks—such as \emph{"Determine whether the image contains text overlays / graphic \& diagram annotations / special marks and output the content-related results"}—and decomposes them into corresponding sub-tasks: \emph{(1) detect text overlays and filter out the topic-related OCR results; (2) detect graphic \& diagram annotations and explain them using the related content; (3) detect special marks and output the results that are not duplicated with previous detection results.} These sub-tasks are formulated in natural language and the corresponding prompt is further refined through human knowledge by providing examples and defining the output formats. Then, these prompts with input images and speech transcipts are passed to the executor based on GPT4o-vision, which performs each sub-task sequentially and finally outputs the detected key frames with topic-related OCR results and content-related explanation (if present). The designed workflow enables the extraction process to generalize across heterogeneous videos while faithfully extracting key frames containing static visual information.

\subsubsection{Dynamic key frames}
To capture camera perspective manipulation, we identify dynamic key frames that represent significant changes in camera viewpoint, such as zoom-ins or close-up transitions. These transitions are widely used in instructional videos to highlight important visual details and draw user attention to critical objects or actions.
We implement this by first applying a scene-based segmentation using PySceneDetect~\cite{PySceneDetect} to obtain coarse scene boundaries. For each segment boundary, we analyze the visual difference between sampled frames before and after the cut. To evaluate whether the frames refer to the same entity or event, we compute a set of perceptual similarity metrics between the two frames, including global and center-cropped structural similarity (SSIM)~\cite{wang2004image}, color histogram distance, ORB feature matching~\cite{rublee2011orb}, and semantic similarity using CLIP~\cite{radford2021learning}.
In addition, we apply monocular depth estimation via a MiDaS~\cite{ranftl2020towards} model to verify whether the transition involves a notable change in scene depth—e.g., a movement from a wide shot to a close-up. When all criteria are met, the timestamp of the boundary is recorded as a dynamic key frame. These points often correspond to visually and semantically meaningful transitions and are later used to enhance note creation. These powerful pre-trained models provide strong guarantees of robustness (\textbf{D4}).

\begin{figure}[t]
    \centering
    \includegraphics[width=0.45\textwidth]{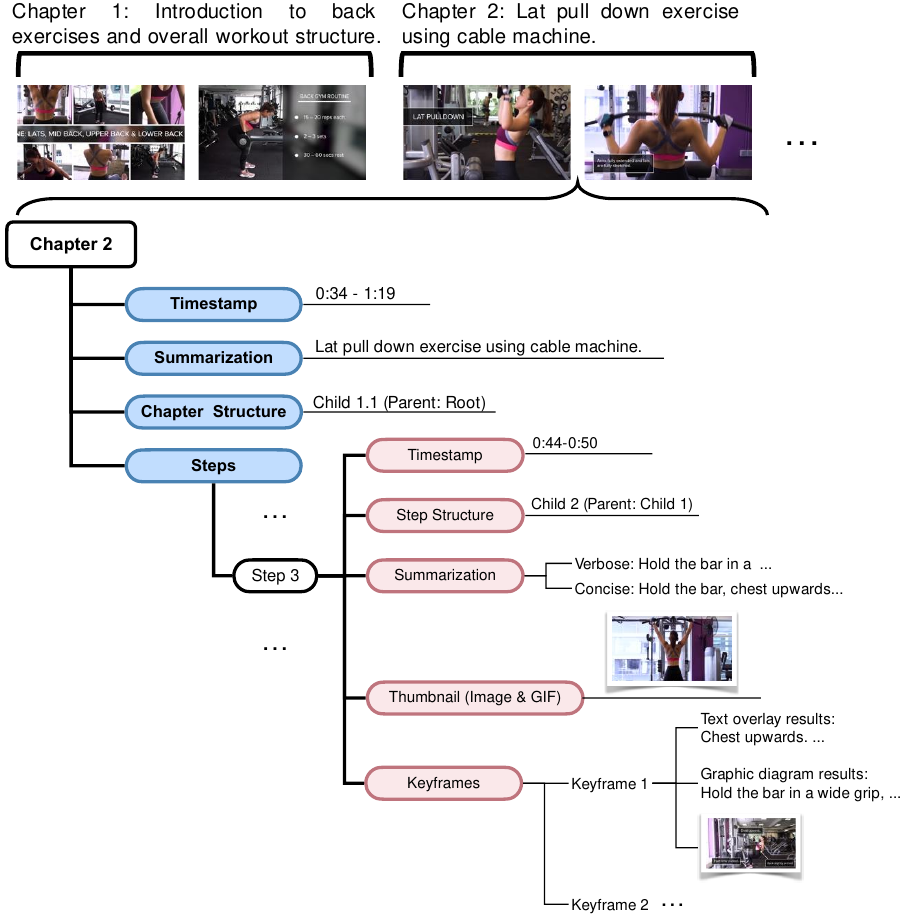}
    \caption{Created note scheme of fitness video for a back gym routine, including structure (chapter and step), summarization (chapter and step), keyframes, thumbnails, and timestamp (chapter and step).}
    \label{fig:notescheme}
\end{figure}

\subsection{Note Creation}
Considering the completeness of the notes, we further create the chapter and step summary and select the thumbnail for each step. Based on the extracted hierarchal structure, visual key information, step summary, and thumbnail, we define a note scheme to represent the note content for the user interface.

\subsubsection{Chapter and step summarization.}
Given the speech transcripts within each chapter, we prompt GPT-4o to generate a concise, high-level summarization encapsulated within a single sentence, allowing viewers to grasp the content quickly. For each step belonging to a chapter, we leverage the corresponding speech transcripts, along with the extracted OCR and graphic \& diagram results of a keyframe (if available) within the step, as the input for step summarization. We then prompt GPT-4o to summarize the input as an instruction step at different levels of detail, including \emph{verbose} and \emph{concise}, associated with the prompt specifying \emph{``summarize a detailed instructional step with some explanations and examples within three sentences and suggest suitable emoji"} and \emph{``summarize a concise instructional step within one sentence"}, respectively (\textbf{D3}). Meanwhile, in order to include the verbal key information, the summarization is also guided by a prompt to \emph{``identify the key information (such as tips and warnings) within the input and highlight them in the summarized step"} (\textbf{D2}).

\subsubsection{Thumbnail retrieval.}
A thumbnail can represent the content of a step. We employ BLIP2 \cite{blip2}, a powerful vision-language model, to retrieve a representative image, which is used as the thumbnail for each step. Given a textual summarization of a step, BLIP2 embeds the text into a shared vision-language space and computes similarity scores against a pre-encoded set of candidate image embeddings, restricted to images falling within the timestamps of the corresponding step. We select the image with the highest similarity as the step thumbnail. For each chapter, we directly use the video frames within the chapter timestamps to generate a GIF to represent the corresponding chapter.

\subsubsection{Note scheme creation and note generation.}

The defined note scheme represents all the extracted information of the instructional video in a clear structure to benefit the following construction of interactable UI. Based on the extracted hierarchical structure and key frames, we incorporate chapter and step summarization, thumbnails, and key frames to the corresponding position according to the index. Given a fitness video for a back gym routine, \oursystem extracts keyframes (text overlays, graphic \& diagram annotations, special marks and camera perspective manipulations), structure (chapter and step), summarization (chapter and step), and thumbnails, and they are integrated into a unique note scheme (Figure \ref{fig:notescheme}), which is used to create the note interface of this fitness video.
We implement the note scheme using predefined HTML \texttt{div} templates with CSS styling and JavaScript interactions, dynamically loading the scheme to generate the interactive notes.

\subsection{User Interface}

\begin{figure}[t]
    \centering
    \includegraphics[width=0.5\textwidth]{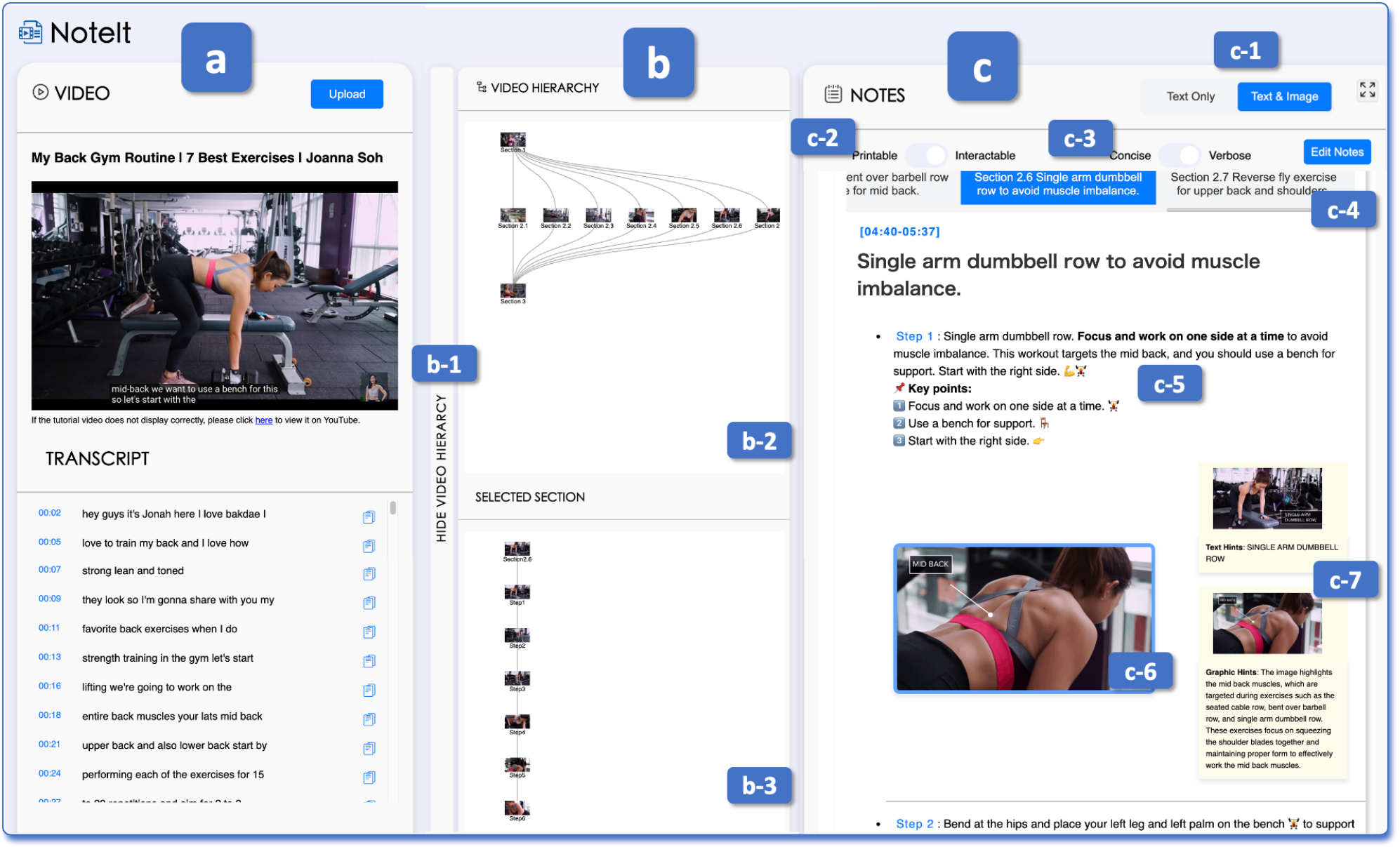}
    \caption{User interface overview. (a) Video player displaying the uploaded video and corresponding transcript. (b) Video hierarchy, with (b-1) a button for displaying or hiding the hierarchy, (b-2) the chapter-level structure, and (b-3) the step-level structure within a selected chapter. (c) Notes, with (c-1) options to customize the modality of the notes (text-only or text with images), (c-2) engagement level (printable or interactable), and (c-3) detail level (concise or verbose). In interactable mode, (c-4) parallel sections are displayed as collapsible slides with tabbed summaries, (c-5) notes for each step, (c-6) a GIF demonstrating the step, and (c-7) text or graphical hints shown in verbose mode (if available).
    }
    \label{fig:interface}
\end{figure}

We developed a web-based interface for \oursystem that allows users to generate, review, and customize notes generated based on the uploaded video (\textbf{D3}). As illustrated in \figurename{~\ref{fig:interface}}, the interface consists of three main components: (a) video player, (b) video hierarchy, and (c) notes. 

\noindent\textbf{\textit{Video Player.}}
The uploaded video and its corresponding transcript are displayed in the left column of the interface (\figurename{~\ref{fig:interface}a}). 

\noindent\textbf{\textit{Video Hierarchy.}} The middle column presents the video hierarchy (\figurename{~\ref{fig:interface}b}), with the upper part (\figurename{~\ref{fig:interface}b-2}) showing the chapter-level structure and the lower part (\figurename{~\ref{fig:interface}b-3}) displaying the step-level structure within a selected section (chapter). Users can select a section by clicking on it in the upper part, and the lower part will update accordingly. Additionally, the middle column can be toggled for visibility by clicking on the botton shown in \figurename{~\ref{fig:interface}b-1}. 

\noindent\textbf{\textit{Notes.}} The right column (\figurename{~\ref{fig:interface}c}) displays the notes generated from the video. Users can customize the note's modality (text-only or text-image, \figurename{~\ref{fig:interface}c-1}), engagement level (printable or interactable, \figurename{~\ref{fig:interface}c-2}), and detail level (concise or verbose, \figurename{~\ref{fig:interface}c-3}). Notes for each step are presented vertically (\figurename{~\ref{fig:interface}c-5}).
In interactable mode, parallel sections are collapsed into slides, and users can navigate through them using a slider with tabs displaying section summaries (\figurename{~\ref{fig:interface}c-4}). In printable mode, both the vertical and horizontal sections are displayed in sequence. Furthermore, in interactable and text-image mode, a GIF is included to demonstrate each step (\figurename{~\ref{fig:interface}c-6}). In verbose mode, the step notes are more detailed, and any text or graphical hints are directly displayed (\figurename{~\ref{fig:interface}c-7}).
Users can also edit the notes or expand them to fill the screen. We will include the video hierarchy editing in the future.

\noindent\textbf{\textit{Interaction Across Components.}} Clicking on a section or step in the video or section hierarchy, the video on the left will jump to the corresponding start time, and the related note will be highlighted in the right column and scrolled to the center of the screen. Additionally, clicking on a time or step index in the note will cause the video to jump to the corresponding start time, ensuring smooth interaction between the video and note sections.

\subsection{Implementation}

\oursystem is implemented with a remote server with one Nvidia GeForce RTX 3090. CLIP uses the pre-trained model \texttt{clip-vit-large-patch 14-336} to encode images for semantic-aware processing, and DINO uses the pre-trainied model \texttt{dinov2\_vitg14} to encode images for vision-aware processing. For speech signal processing, \oursystem uses OpenAI \texttt{Whisper-large} for speech transcription and sentence segmentation. Moreover, \oursystem is powered by OpenAI \texttt{GTP-4o (GPT4o-vision)} API for captioning and reasoning. BLIP2 use \texttt{blip2 -itm-vit-g} to encode text and images as text embeddings and vision embeddings. The processing time of NoteIt is sensitive to server performance, for example, a 7:10 video takes about 25 minutes to process. Computation cost can be optimized with advanced MLLMs’ API or keyframes clustering to narrow the extraction space.
The user interface is implemented using D3.js for interactive data visualization, with Python and the Flask framework serving as the backend to manage server-side logic and API integration. Additionally, the system leverages the YouTube API to retrieve video metadata and content.

\section{Technical Evaluation}
\label{technicaleval}
To demonstrate the effectiveness and generality of \oursystem, we conducted a technical evaluation by measuring the performance of hierarchical structure extraction and the performance of visual key information extraction.

\subsection{Datasets}
We selected 32 instructional videos in 8 categories from the original 80 videos we analyzed in Section \ref{designspace}. We selected 4 videos for each category to diversify the video categories, and we also considered the following criteria: (1) the video clearly shows the hierarchical structure, with both vertical and horizontal structures; (2) key information are presented both verbally and visually; (3) the video contain at least two of the visual key information (text overlay, graphic \& diagram annotations, special marks and camera perspective manipulations), and the more the better. The average length of the selected videos was 8.63 minutes (min = 3.91 minutes, max = 15.27 minutes). The purposefully chosen maintain the same level of diversity and rich characteristics, ensuring the evaluation reflects a broad range of real-world conditions.

\subsection{Method}

\oursystem parses all the 32 instructional videos to generate corresponding notes. To evaluate the extraction capabilities of hierarchical structure and key frames of visual key information, we asked 10 in-house raters from our university to label all 32 instructional videos, which serve as the ground truth. All raters were trained by the annotation process, and they were blinded to our evaluation method. Each video was annotated by 2 people and further checked by 2 raters.
The raters annotate the chapter structure and frames containing both static (text overlays, graphic and diagram annotations, and special marks) and dynamic key information (camera perspective manipulations).

For hierarchical structure, we mainly consider the structural segmentation instead of the time boundary because it has an objective evaluation criterion.
Although segmenting a video into chapters and steps is subjective \cite{fraser2020temporal}, reaching a consensus is relatively easier for chapter-level segmentation compared to step-level segmentation. Therefore, our evaluation of the hierarchical structure focused primarily on chapter-level segmentation. To assess it, we used Mean Relative Accuracy (MRA) \cite{yang2024thinking} in the range of 0 to 1, which measures how well the chapter-level structure extracted by \oursystem aligns with the ground truth. A higher MRA score closer to 1 indicates better alignment and correctness in chapter segmentation.
For key information extraction capability, we calculated Recall, Precision, and F1-score, the widely used metrics \cite{tutoai,Soloist,SlideGestalt}, for evaluation.
Each metric in the range between 0 and 1 is computed by comparing the frames annotated by the raters (ground truth) with those extracted by \oursystem. 

\subsection{Results}

\begin{table}[t!]
    \centering    
    \small
    \caption{Visual key information extraction performance.}
    \begin{tabular}{c|ccc}
        \toprule
         & Avg. Recall  & Avg. Precision  & Avg. F1 \\
        \midrule
        \makecell[c]{Text overlays, \\Graphic \& Diagram \\annotations, \\Special marks} & 91.88\% & 92.20\% & 91.63\% \\
        \midrule
        \makecell[c]{Camera perspective\\ manipulation} & 67.94\% & 90.96\% & 70.86\% \\
        \midrule
        Average & 79.91\% & 91.58\% & 81.24\% \\
        \bottomrule
    \end{tabular}
    \label{tab:three_line}
\end{table}

\begin{figure*}
    \centering
    \includegraphics[width=\linewidth]{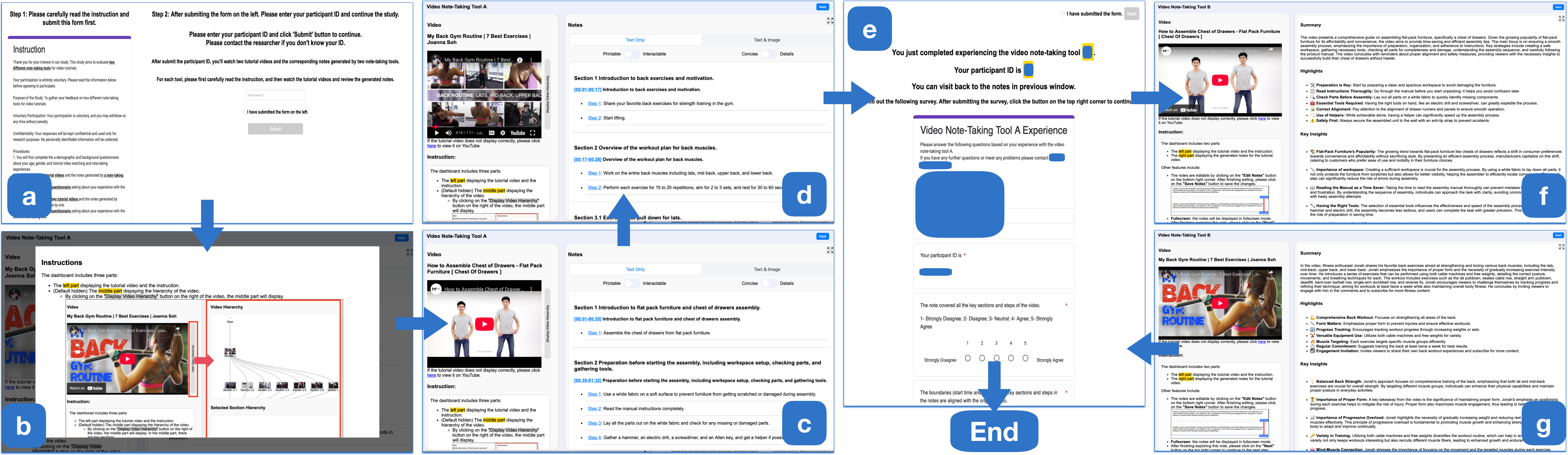}
    \caption{Overview of the user study procedure. The user first lands on the (a) welcome page. Upon entering each tool page, an instruction screen (b) is shown. After clicking the confirm button at the end of the instruction, the instructional video and corresponding note are displayed (c). Once the first video is completed, the user proceeds to another video-note pair generated by the same tool (d). After reviewing both videos, the user completes a survey evaluating the first tool (e), then moves on to review notes generated by the second tool (f, g). A final survey (e) follows to assess the second tool.}
    \label{fig:procedure}
\end{figure*}

\tablename{~\ref{tab:three_line}} presents static and dynamic visual key information extraction performance of \oursystem. 
For static scenarios including text overlays, graphic and diagram annotations, and special marks, \oursystem successfully captures nearly all key infomation with minimal false positives, as evidenced by consistently high precision, recall, and F1 scores all above 91\%. 
In contrast, dynamic scenarios, characterized by camera perspective manipulation, involve more implicit and less standardized patterns, often lacking stable visual markers. 
Though challenging, \oursystem achieves a high precision of 90.96\%, indicating that most of the dynamic visual key information it identifies is indeed relevant and important. 
The recall of 67.94\% suggests that \oursystem effectively identifies prominent and noticeable dynamic cues, while more subtle or less salient instances may be missed due to the implicitness and variability in dynamic visual content. 
For example, overly gradual camera transitions might produce imperceptible changes between frames, hindering the identification of dynamic cues. Additionally, scenarios involving excessively rapid zoom-ins that lose surrounding context make it difficult to discern whether consecutive frames are part of the same overall scene. 
Notably, some frames in camera perspective manipulation may not necessarily contain information crucial for task progression. Since our entire pipeline is designed to help users learn from instructional videos, missing certain subtle or less noticeable frames does not impact its overall effectiveness. A recall of 67.94\% is sufficient to capture the most critical dynamic visual cues, providing meaningful support to users without compromising the learning process. Furthermore, we discuss this potential improvement in the Limitations section. 
Overall, \oursystem effectively and reliably extracts the majority of both static and dynamic visual key information.

To assess chapter-level segmentation, we calculate the average MRA score across all 32 instructional videos. \oursystem achieves an average MRA score of 75.31\%, with 11 videos exhibiting a perfect MRA score of 1.0, indicating that all chapters in those videos are correctly identified. These results underscore \oursystem’s strong alignment with human interpretations, demonstrating the model's advanced capability in parsing complex video narratives.

\section{User Study}

We conducted a user study to assess the feasibility of our pipeline and the overall usability of the user interface. The same 32 videos as in the technical evaluation (Section \ref{technicaleval}) are used for the user study.

\subsection{Participants}

We used G$^*$Power \cite{faul2009statistical} to determine the required sample size for a Wilcoxon Signed-rank test \cite{rosner2006wilcoxon} with a moderate effect size ($d=0.5$), significant level $\alpha=0.05$, and power $(1-\beta)=0.8$, resulting in a minimum of 35 participants. Accordingly, we recruited 36 participants via word-of-mouth and online advertisement in our university community. All the participants aged between 22 and 44 ($M=27.1$, $SD=3.9$), including 25 self-identified males and 11 females. One participant was a tutorial video creator, and three had authored tutorial notes. All had prior experience with watching or creating instructional videos and note-taking. None of the participants had experienced our system or the baseline system before they came. Informed consent was obtained, and each participant received a \$10 gift card for participation.

\subsection{Procedure}

We adopted a within-subject design in which each participant evaluated notes generated by both \oursystem and a baseline LLM-based note generation commercial application, NoteGPT \cite{notegpt}.
NoteGPT supports functionalities such as AI presentation, AI homework helpler, and AI book library. However, they are out of this paper's research scope. We mainly use its note generation function of converting YouTube videos into notes, which mostly aligns with this paper's research scope, making it an appropriate point of comparison for evaluating our approach.
The study was conducted on a dedicated web-based platform (\figurename{~\ref{fig:procedure}}), which integrated both tools to ensure a consistent and autonomous user experience. Specifically, we copied the outputs generated by NoteGPT for our study videos into the platform for comparison. 

Participants accessed the study via a shared link distributed through email and messaging platforms. 
Upon landing on the welcome page (\figurename{~\ref{fig:procedure}a}), participants received study instructions, provided consent, and proceeded to review two instructional videos with corresponding notes from the first tool (\figurename{~\ref{fig:procedure}d,c}). They then completed a questionnaire assessing their experience (\figurename{~\ref{fig:procedure}e}). This process was repeated for the second tool using the same two videos (\figurename{~\ref{fig:procedure}f,g}).
To preserve anonymity during the study, we referred to the two tools as Tool A (\oursystem) and Tool B (the baseline tool) throughout all participant-facing materials and analysis. 
We also employed a counterbalanced study design to control for order effects: half of the participants used Tool A first, followed by Tool B, while the other half followed the reverse order. This ensured that any observed differences in user experience or performance were not biased by the sequence in which the systems were used.
Each post-tool survey included 5-point Likert-scale items (1 strongly disagree - 5 strongly agree) \cite{jiang2022understanding} evaluating consistency, informativeness, adaptability, and overall satisfaction, along with open-ended feedback. 
For \oursystem, additional questions assessed the note customization features.
We also conducted the System Usability Scale (SUS) \cite{bangor2008empirical} study to evaluate the overall usability of the entire system.

\subsection{Results}
The results are reported in \tablename{~\ref{tab:result}}, including the mean and standard error of users' ratings. For questions applicable to both \oursystem and baseline, we performed two-sided Wilcoxon signed-rank tests to compare user ratings. The $p$ values less than 0.05 were considered statistically significant.
As for the overall system usability, \oursystem achieved an average SUS score of 78.1 ($SE=2.02$, $SD=12.11$).

  \begin{table*}[!t]
      \small
      \renewcommand{\arraystretch}{1}
        \caption{The result of statistic analysis. We report the mean $M$ and standard error $SE$ for the baseline tool and \oursystem, and the test statistic $W$ and $p$ value of the Wilcoxon signed-rank test. All tests were two-sided, where p-value < 0.05 indicates the statistical significance. }
        \label{tab:result}
        \begin{tabular}{lccccccccccccc}    
          \toprule
           \bfseries &  \multicolumn{2}{c} {$M$ ($SE$)}&  & \\
           \cmidrule{2-3}
          \bfseries &  Baseline & \oursystem & $W$ & $p$ \\
          \midrule
          Consistency: Notes cover all key sections and steps in the video. (Q1) & 3.39 (0.21) & 4.78 (0.07) & 5.5 & <0.001$^*$\\
          
          Consistency: Key section and step boundaries in notes align with the video. (Q2) & 2.61 (0.23) & 4.58 (0.09) & 4.5 & <0.001$^*$\\
          Consistency: Notes followed the video structure with correct step order and clear parallel steps. (Q3) & 2.69 (0.22) & 4.72 (0.09) & 9.0 & <0.001$^*$\\
          
          Informativeness: Notes capture key verbal information in the video. (Q4) & 3.03 (0.21) & 4.64 (0.08) & 0.0 &  <0.001$^*$ \\
          Informativeness: Notes capture key visual information in the video. (Q5) & 2.28 (0.20) & 4.44 (0.10) & 3.0 & <0.001$^*$\\
          Adaptability: Notes support diverse video categories and characteristics. (Q10) & 2.92 (0.21) & 4.44 (0.11) & 18.0 & <0.001$^*$\\
          Overall Satisfaction: Users prefer using the generated notes over rewatching the video. (Q11)  & 2.75 (0.23) & 4.31 (0.15) & 5.0 & <0.001$^*$\\
          Customization: The necessity of customizing note detail levels. (Q6)  & - & 4.33 (0.13) & - & - \\
          Customization: The necessity of customizing engagement (printable or interactable). (Q7)  & - & 4.17 (0.14) & - & - \\
          Customization: The necessity of customizing note modality. (Q8)  & - & 4.50 (0.13) & - & - \\
          Customization: The satisfaction after customization. (Q9)  & - & 4.50 (0.09) & - & - \\
          UI Design Usefulness: The usefulness of structure inspection via tree. (Q12) & - & 4.33 (0.13) & - & -\\
          UI Design Usefulness: The usefulness of locating content via tree click. (Q13) & - & 4.72 (0.08) & - & -\\
          UI Design Usefulness: The usefulness of switching note modalities (text only and text \& image). (Q14) & - & 4.33 (0.15) & - & -\\
          UI Design Usefulness: The usefulness of switching detail levels (concise and detailed). (Q15) & - & 4.31 (0.17) & - & -\\
          UI Design Usefulness: The usefulness of switching engagement modes (printable and interactable). (Q16) & - & 4.00 (0.17) & - & -\\
          \bottomrule
        \end{tabular}
    \end{table*}

\subsection{Findings}
Participants rated \oursystem significantly higher than the baseline across all design goals and evaluation metrics.

\noindent\textit{\textbf{\oursystem provides consistent hierarchical structure aligned with the video. 
}}
\oursystem achieved the highest average score of 4.69 for Consistency, significantly outperforming the baseline ($p$<0.001).
Participants noted that its hierarchical segmentation, clear step labeling, and text-image pairing helped them follow the video more logically and retain the sequence of actions (\textbf{D1}). Many appreciated how the notes mirrored the video structure, reducing the need to rewatch. These findings align with the previous research that structure facilitates video comprehension \cite{Subgoal,RecipeScape}.
For example, P2 stated, ``The detailed steps provided can easily guide me to understand what I should do next,'' and P29 remarked on the efficiency of navigating the restructured content. Only one participant (P34) reported a missing step, and P34 raised the question ``Can I add the missing step in the system?"
In contrast, while some users acknowledged that the baseline tool helped convey the video's general idea (P7, P23), 23 out of 36 participants reported missing key steps. The baseline performed reasonably well on simple-structured videos but struggled with longer or more complex content. In such cases, \oursystem consistently captured more complete and accurate instructional sequences.

\noindent\textit{\textbf{\oursystem captures key verbal and visual information in the video.}}
\oursystem significantly outperformed the baseline in conveying key verbal and visual cues.
Most participants found the notes informative and concise, with 14 out of 36 specifically praising the combination of text and images or GIFs for enhancing comprehension (\textbf{D2}). The system's ability to summarize key steps, visuals, and timelines helped users focus on essential content while skipping redundant parts. As P5 noted, ``It provided structural, vivid, and illustrative notes... which helped me better understand and memorize the key content of the video.''

While a few users pointed out minor omissions, such as specific quantities in cooking videos (P33), the baseline tool showed more frequent gaps, missing ingredients (P18), numerical values (P16), and instructional steps (P8). P33 also suggested ``It would be better if the system could allow me to define the visual information that I think is important in different videos." A few participants also noted that some visual outputs from \oursystem were overly detailed or redundant (P29).

These findings align with prior studies that suggested the importance of visual and verbal key information \cite{VideoSticker,HowToCut,chi2013democut}. They also demonstrate that \oursystem not only improves information density and clarity but also offers a more structured and memory-friendly experience. While occasional gaps remain, the overall enhancement in note informativeness and user comprehension reflects the potential of multimodal note-taking systems in supporting diverse video-based learning tasks.

\noindent\textit{\textbf{\oursystem provides necessary and useful customization.}}
\oursystem received an average score of 4.38 for customization. 
P5 described the interactive mode as ``surprisingly good and illustrative,'' while P28 appreciated choosing modes based on their needs. 
Most participants (29 out of 36) (strongly) agreed that the ability to switch between concise and detailed notes was useful, though some, like P3, found the detailed version unnecessary for simpler videos. 
A few participants (e.g., P17, P22, P34) criticized excessive emojis and detailed text hints for adding cognitive load.

For engagement modes (printable vs. interactable), most users supported the feature, though some found it less distinguishable in simpler scenarios where differences between modes were minimal (P6) and always preferred the interactable version (P7, P19, P21).
Regarding modality customization (text-only vs. image/GIF-text pairs), the majority of participants (32 out of 36) supported the feature, while P2 and P17 described the text-only mode as redundant.
Additionally, P23 noted that while customization was helpful, too many options could become overwhelming.

We also asked the participants to rate the usefulness of UI components in \oursystem, such as inspecting and navigating the tree structure and selecting presentation modes. All components scored no less than 4.0, with ``locating notes from the tree'' receiving the highest rating ($M=4.72$, $SD=0.45$) and the selection between printable and interactable modes the lowest ($M=4.0$, $SD=1.01$). 

These results indicate that users valued the system's customization to personal learning styles and task complexity \textbf{(D3)}, which aligns with what we analyzed in Section \ref{charnotes}. Visually rich, interactive, and concise formats were generally preferred. However, effective UI design should balance flexibility with simplicity to avoid unnecessary cognitive burden \cite{jiang2023healthprism}.

\noindent\textit{\textbf{\oursystem supports diverse video categories and content.}}
\oursystem was rated significantly higher than the baseline in adapting to instructional videos of varying length, structure, and complexity ($p$<0.001). 
The participants found the baseline tool was able to provide summaries and identify some key points for relatively shorter instructional videos with a simple structure (e.g., a first-aid video without horizontal structure and a make-up video with only one horizontal alternative, P29, P30). However, it struggled with more complex content (e.g., a fitness video with nine alternative parallel sections and a cooking video with details about ingredient amounts, P10, P18).
In contrast, \oursystem consistently handled diverse structures and topics more effectively, earning higher ratings across video types (\textbf{D4}). While a few participants (e.g., P15, P23) were curious about whether other videos could be processed.

\noindent\textit{\textbf{High satisfaction of \oursystem.}}
\oursystem received significantly higher ratings for overall satisfaction than the baseline ($p$<0.001). Participants particularly appreciated its clear structure (15/36), efficiency (21/36), ease of learning and reflection (13/36), and multimodal presentation (8/36).
Many noted they could quickly access relevant content and skip unnecessary parts. As P6 stated, ``It provided me with a clear structure of the video and I can easily find the note for a specific step,'' and P29 highlighted its efficiency in helping users locate desired segments.
P17 remarked, ``I can understand a 6-min video by looking at the notes within 1 minute,'' but criticized the baseline notes as poorly structured and hard to follow.

\noindent\textit{\textbf{High usability of \oursystem.}}
\oursystem achieved an average SUS score of 78.1 ($SE=2.02$, $SD=12.11$). The lowest score is 57.5 (P35), and the highest score is 100 (P9 and P25). 
Most users found the tool intuitive, with features like note-to-video linking, quick navigation, and multi-modal displays improving usability. The ability to jump to video sections, review and reflect without rewatching, and track progress contributed to an overall efficient and accessible user experience.
``I can skip the parts I already know, and I can understand a video by just reading the notes.'' (P17)
``It helped me to quickly find the part I am interested in.'' (P23)
A few participants pointed out that it required some effort to learn how to use this tool (P15, P35, P36).

\subsection{Discussion}
Despite high satisfaction, we reflect on the findings and participants' suggestions to discuss design opportunities to improve \oursystem.

\noindent\textit{\textbf{Generalize to diverse instructional videos.}}
Participants suggested that our system could support user-defined visual information extraction (P33) and note generation for various videos (P15, P23). Although NoteIt are capable of processing other videos beyond the selected videos, it is important to enhance the framework to adapt to novel input features and thus generalize to different videos. As existing visual extraction module is powered by foundation models with zero-shot ability \cite{clip,GPT4o}, we suggest that future research carefully design the sub-module based on MLLM and foundation models in NoteIt and integrate this sub-module into NoteIt to adaptively extract the user-defined key frames. Given this adaptation for novel input, it is possible to extend the existing framework to process different videos with distinct features.

\noindent\textit{\textbf{Customization Enhancement.}}
Participants recommended enhancing customization by supporting figure editing (P6), handwriting input (P10), sticky headers for easier navigation (P31), a manual highlight feature (P15), and timestamp and GIF edits (P33). P23 pointed out that too many options could overwhelm users. 
These insights highlight the need for providing flexible yet intuitive customization features that align with users' individual preferences and cognitive strategies \cite{jiang2025dietglance}.
Rather than offering exhaustive configuration, effective customization should balance control with simplicity, enabling users to tailor the experience without becoming burdened by interface complexity. This suggests a direction for supporting more adaptable interfaces that scale with user expertise and task demands.

\noindent\textit{\textbf{Advanced Interaction.}}
While participants can edit and save the generated notes, they recommended enhancing interactivity by supporting prompt-based note regeneration and question answering (P16, P23, P28, P33) and revisable hierarchical structure (P12). In response to participants' expectations, we suggested that future work could explore responsive and intelligent modes of user interaction.
Inspired by recent works that enhance interaction with LLM \cite{peng2025navigating}, we look forward to investigating the potential of integrating LLM-powered conversational agents and context-aware interfaces to support real-time clarification, retrieval, and personalization, transforming note-taking into a more interactive, learner-centric dialogue. In addition, visualizing hierarchical structure in NoteIt is a novel way to enhance user understanding of instructional tasks. For complex tasks in particular, a clear or well-organized structure could possibly improve learning efficiency. Therefore, it is important to allow users to revise or control the hierarchical structure according to their understanding and learning preferences.

\noindent\textit{\textbf{Integrate with other tools.}}
Participants emphasized the value of integrating with other tools, such as mind maps (P19), Notion (P21, P22), and generating flowcharts or diagrams from video content (P17). These suggestions reflect a need for seamless workflow integration across platforms. For example, studies have shown that digital mind mapping tools can enhance academic performance by improving comprehension and organization of complex information \cite{hazaymeh2022effectiveness}. Knowledge Graphs offer a promising approach to represent complex relationships among different steps and concepts \cite{hogan2021knowledge}. 
We look forward to integrating visual expressiveness and inclusive design for more effective information management.

\noindent\textit{\textbf{Visual and Accessibility Improvements.}}
Participants recommended enhancing visual clarity by distinguishing key details (P24), using more representative images or generated icons to better summarize sections (P29), and adding a colorblind mode to enhance accessibility (P32). 
These recommendations highlight the critical role of intuitive and inclusive visual design in instructional systems. Prior research has shown that the effective use of visual differentiation, representative imagery, and accessibility features not only improves content comprehension but also significantly elevates overall user engagement and satisfaction \cite{seifi2024influence}. 
Additionally, aesthetic manipulation can influence users' perceived usability \cite{tuch2012beautiful}. We suggest integrating well-considered visual and accessibility features to enhance user engagement and satisfaction.

\noindent\textit{\textbf{Summarization and Metadata Support.}}
Participants emphasized the need for clearer summarization and metadata. Suggestions included adding a summary or video overview (P30, P35), showing total and segment durations (P11), and providing extended advice beyond the original video (P20). These suggestions highlight the value of structured, contextual guidance in facilitating efficient knowledge acquisition and comprehension. Integrating concise summaries and comprehensive metadata can help learners quickly have an overview of the video content and assess its relevance. We look forward to augmenting existing video note-generating tools like NoteGPT \cite{notegpt} and NotebookLM \cite{notebooklm} to provide more fine-grained key information, structured navigation, and supplemental guidance, ultimately enhancing the user's ability to effectively comprehend and engage with instructional videos.
\section{Limitation And Future Work}

In this section, we discuss the limitations of \oursystem, and propose future directions that could further advance research in the field of automatic note generation.

\textbf{\textit{Beyond instructional videos and learning notes.}} 
As mentioned in the paper, our study centers on instructional videos; accordingly, we only select a certain number of instructional videos and notes for design space analysis based on specific criteria. Given the impracticality of analyzing every instructional video and notes, we adopt purposive sampling to select representative instructional videos and derive common patterns from them, which has been proven to be a feasible solution in the HCI community \cite{tutoai,chi2013democut,HowToCut}. The selection criteria, spanning diverse categories, creators, tasks, and durations, ensure the representativeness of the selected videos. Therefore, the derived design goals is reliable and generalizable, despite being derived from a limited set of instructional videos. We believe that, grounded in these design goals, NoteIt can generalize well beyond the selected videos to wide range of instructional videos.
While other video types were considered outside the scope, \oursystem exhibits strong generalizability, suggesting the potential of our pipeline to be extended beyond its current intent. 
We envision future research directions of expanding the video-to-note conversion in broader domains, such as entertainment and casual contexts—covering video types like game streaming and vlogs—as well as supporting alternative note formats, including flyers and posters.

\textbf{\textit{Unsatisfactory outcomes in video processing.}} 
While our technical evaluation indicated strong performance in extracting visual key information across the eight categories examined, we observed some missing key frames of visual information in a small number of videos. One contributing factor is camera perspective manipulation: extremely rapid shot transitions can disrupt entity and event alignment between consecutive frames, whereas extremely gradual zoom-ins or pans may not be recognized as distinct scene cut. Another challenge arises from specialized instructional formats, such as multi-window or slide-like layouts (e.g., displaying multiple sub-frames or simultaneously overlaying multiple views). These layouts can interfere with or fragment the main content, thereby undermining depth estimation, CLIP similarity, and feature matching methods, which assume a single coherent scene per frame. A potential solution could be to integrate advanced video encoders~\cite{VTimeLLM,zhang2023video} and video processing tools~\cite{ravi2024sam} that better capture overall scene context and user-relevant actions, subsequently incorporating those outputs into cross-frame similarity assessments.

\textbf{\textit{Form factors of the generated notes.}} On the note generation side, we support a series of presentations that have been commonly used in online textual media posts. Meanwhile, we support users printing out the notes for further offline references. Yet, the supported formats and layouts are pre-defined. Each video has its distinct features and in some cases, it may not be appropriate to classify them into a fixed set of formats or layouts. In future work, we could consider the adaptive presentation formats and layouts that respond to the specific features of each video. One promising direction is to further unleash capabilities of LLM in design \cite{Graphimind,LayoutLLM}, where LLM can generate the HTML code for website formats and layout according to the videos summarization. Nevertheless, integrating such adaptability into our current MLLM-based pipeline will require significant effort and further exploration.

\textbf{\textit{Expertise-Aware Note Customization}}. While \oursystem allows users to select the detail level of notes and jump to the notes of interest through the video hierarchy, the note is not currently directly adapted to the individual expertise of users. For example, a user who is proficient in foundation makeup but lacks experience with eye makeup may prefer that the notes for a makeup tutorial provide a simplified description or even omit details on foundation application, while offering more detailed guidance on eye makeup techniques. 
This highlights the need for expertise-aware customization that adjusts the notes based on a user's specific proficiency, ensuring that the note content is both relevant and efficient for users at varying skill levels.
To address this, we could allow users to provide key information about their expertise and expectations regarding the notes before video processing. This information could then be incorporated into the processing pipeline to generate more customized notes tailored to individual needs. 
Such a feature would enable \oursystem to offer a more personalized learning experience, enhancing user engagement and knowledge retention across diverse levels of expertise.

\textbf{\textit{Advanced notes system capabilities.}}
Our current pipeline extracts visual key information primarily from frames featuring text overlays, visual cues, and camera perspective manipulations.  Scene coverage can be further broadened by exploiting video frames aligned with salient linguistic markers in the instructor's narration, such as deictic expressions (e.g., “here” or “this way”), syntactic cues that foreground importance, and discourse-level references that link prerequisite actions to later steps. The potential solution is to incorporate an LLM-centric agent \cite{VideoAgent} to understand and capture the above semantic information and decompose it into sub-goals to extract corresponding frames, comprehensively covering salient scenes.
NoteIt purposely centers on the formats of interactable notes because notes further augment the prevailing mixed-media tutorial paradigm by providing flexible presentation without relying on video as a basis and integrating fluent multimodal information. This facilitates more effective comprehension of instructional video content, thereby constructing knowledge. To further improve the interaction ability, we can integrate audio guidance and conversational walkthrough for the hands-busy settings (e.g., cooking and repair). To implement it, we could export the generated notes to speech via existing text-to-speech techniques \cite{SeedTTS} to narrate the content or further integrate it with an LLM-powered Retrieval Augmented Generation (RAG) module to enable conversation \cite{ragaudio}.

\textbf{\textit{Other potential applications.}} Besides note generation for instructional videos, NoteIt offers significant potential for diverse applications, such as collaborative sessions, entertainment media analysis, and journalism.
For example, structured multimodal notes of collaborative meetings or workshops could help teams quickly revisit critical discussions and outcomes, enhancing documentation efficiency and asynchronous collaboration. Additionally, NoteIt can generate concise summaries of films and TV shows, highlighting key plot points, themes, and character developments, or create structured notes from news events, press conferences, and interviews, facilitating rapid content review and analysis.

\section{Conclusion}

In this paper, we presented \oursystem, a novel system for converting instructional videos into structured, user-customizable notes. 
We began by analyzing the unique characteristics of instructional videos and the diverse needs of note consumers, which informed the development of a comprehensive design space for note generation. 
Based on this understanding, we introduced an end-to-end pipeline that extracts hierarchical structure and key multimodal information from instructional videos to generate high-fidelity notes. 
To support practical usage, we also designed an interactive user interface that allows users to upload videos and tailor the note format according to their preferences.
To evaluate the effectiveness of our system, we conducted both a technical evaluation and a comparative user study. The results from these complementary evaluations demonstrate that \oursystem performs effectively in both content accuracy and user satisfaction.
Together with the identified limitations and future directions, we believe this work opens up a promising line of research in automatically converting instructional videos into structured, adaptable learning materials.

\begin{acks}
We gratefully acknowledge the authors, Xinchen Zhang, Chirui Chang, and Handi Chen, for their invaluable and equal contributions to this work. We thank the reviewers and editors for their constructive suggestions. This work was supported by the UGC General Research Fund no. 17209822 and the Innovation and Technology Commission Fund no. ITS/383/23FP from Hong Kong.
\end{acks}

\bibliographystyle{ACM-Reference-Format}
\bibliography{99_references}

\appendix

\end{document}